\documentclass[journal]{IEEEtran}
\usepackage{amsmath,amsfonts,amsthm,mathtools}
\usepackage{algorithmic}
\usepackage{array}
\usepackage[caption=false,font=normalsize,labelfont=sf,textfont=sf]{subfig}
\usepackage{textcomp}
\usepackage{stfloats}
\usepackage{url}
\usepackage{verbatim}
\usepackage{graphicx}
\usepackage{xcolor}
\usepackage{nomencl}
\usepackage{etoolbox}
\usepackage{ifthen}
\usepackage{booktabs}
\usepackage{adjustbox}
\usepackage[version=4]{mhchem}
\makenomenclature
\usepackage{cite}
\usepackage{multirow}
\usepackage{bm}
\usepackage{makecell}
\usepackage{enumitem}
\usepackage{caption}
\allowdisplaybreaks 
\usepackage[subtle,tracking=normal]{savetrees}

\def\BibTeX{{\rm B\kern-.05em{\sc i\kern-.025em b}\kern-.08em
    T\kern-.1667em\lower.7ex\hbox{E}\kern-.125emX}}
\usepackage{balance}
\usepackage{algorithm}

\usepackage{url}
\usepackage{hyperref} 
\usepackage{xcolor}
\usepackage{pifont}

\hypersetup{
    colorlinks=true, 
    linkcolor=blue, 
    filecolor=black, 
    urlcolor=blue, 
    citecolor=blue, 
    linkbordercolor={1 1 1}, 
    pdfborder={0 0 0} 
}

\begin{document}

\bstctlcite{IEEEexample:BSTcontrol}
\title{A Process-Aware Demand Response Evaluation \\ Framework 
for Hydrogen-Integrated Zero-Carbon  \\Steel Plants Coupled with Methanol Production}

\author{\IEEEauthorblockN{Qiang Ji, \textit{Student Member, IEEE}}, \IEEEauthorblockN{Lin Cheng, \textit{Senior Member, IEEE}},
\IEEEauthorblockN{Yue Zhou, \textit{Member, IEEE}},\\
\IEEEauthorblockN{Ning Qi, \textit{Member, IEEE}},
\IEEEauthorblockN{Kaidi Huang, \textit{Student Member, IEEE}},
\IEEEauthorblockN{Jianzhong Wu, \textit{Fellow, IEEE}},
\IEEEauthorblockN{Ming Cheng},
 \thanks{Manuscript created xxx, 2026; revised xxxx, 2026; revised xxxx, 2026; accepted xxxx 2026.
 This work was supported by Smart-Grid National Science and Technology Major Project (No. 2025ZD0806300). (\textit{Corresponding author}: Yue Zhou.)
 
 Qiang Ji,  Lin Cheng, and Kaidi Huang are with the Department of Electrical Engineering, Tsinghua University, Beijing 100084, China (e-mail: jiq23@mails.tsinghua.edu.cn).\par
Yue Zhou is with the State Key Laboratory of Smart Power Distribution Equipment and System, Tianjin University, Tianjin, China. (e-mail:
zhouyue68@tju.edu.cn )\par
Ning Qi is with the Department of Earth and Environmental Engineering, Columbia University, New York, NY
10027 USA (e-mail: nq2176@columbia.edu).\par
Jianzhong Wu is with the School of Engineering, Cardiff University, Cardiff, CF243AA, UK (e-mail: WuJ5@cardiff.ac.uk).\par
Ming Cheng is with the Department of Science and Technology,
State Grid Jibei Electric Power Co., Ltd., Beijing, China. (e-mail:
Ericcheng0518@163.com).
}
}

\markboth{IEEE TRANSACTIONS ON Smart Grid,~Vol.~X, No.~X, XX August~2026}
{How to Use the IEEEtran \LaTeX \ Templates}

\maketitle

\begin{abstract}
High penetration of intermittent renewables (RES) and the retirement of thermal units have aggravated the flexibility scarcity in power systems. Hydrogen-based low-carbon steel production systems possess substantial demand response (DR) potential. This paper proposes a process-aware DR evaluation framework for hydrogen-integrated zero-carbon steel plants coupled with methanol production (\ce{H2}-DRI-EAF-MeOH). First, a novel \ce{H2}-DRI-EAF-MeOH production architecture is proposed to eliminate residual emissions via methanol synthesis, where the integrated energy-material flows are formulated to reflect the complex coupling interactions governing system DR potential. Second, to capture electric arc furnace (EAF) operational constraints while preserving optimization tractability, an operating feasible region model is developed and validated using field data from a pure hydrogen direct reduced iron and EAF plant, yielding an average relative error of 4.1\%. Third, a process-aware DR potential evaluation model is proposed, incorporating a nonlinear asymmetric penalty formulation and an adaptive rolling mechanism to reflect operators’ inherent aversion to process deviations and avoid myopic scheduling. Finally, dual-side DR potential evaluation metrics are established to quantify grid-side delivered DR capacity and ramping risks, while making load-side unit-level regulation behaviors observable and measurable. Case studies demonstrate that the proposed process-aware DR evaluation model achieves an average effective delivered DR capacity of 178.3 MW, improves the RES-load matching degree from 0.257 to 0.587, and reduces operational costs by 15.68\% compared with the baseline scheme. Furthermore, the proposed exponential asymmetric penalty model mitigates extreme tail risks of process deviations. Ultimately, the proposed framework provides a theoretical foundation for leveraging RES-steel-chemical synergies to mitigate flexibility scarcity.
\end{abstract}

\begin{IEEEkeywords}
Demand response, renewable energy integration, hydrogen-based steelmaking, electric arc furnace, methanol production, industrial flexibility
\end{IEEEkeywords}

\section{Introduction}\label{Introduction}
\IEEEPARstart{D}{riven} by climate mitigation goals \cite{lei2023global}, the growing penetration of intermittent renewables (RES) \cite{Yan1} and the decommissioning of conventional thermal power units have created a critical flexibility deficit, which triggers substantial RES curtailment and undermines system stability \cite{Lai1}. Demand response (DR) \cite{palensky2011demand}, implemented through price-based or incentive-based mechanisms, enables users to adjust their electricity consumption patterns, thereby enhancing power system flexibility and facilitating RES integration \cite{yu2023demand}. Among various types of electricity consumers, industrial loads are regarded as the most promising DR resources \cite{shoreh2016survey,golmohamadi2022demand}. With the transition toward low-carbon manufacturing, industrial loads exhibit higher electrification levels and greater responsiveness to price signals \cite{wei2019electrification}. Therefore, quantifying and unlocking the DR potential of these low-carbon industrial loads has become increasingly important for enhancing power system flexibility and facilitating large-scale RES integration. \par
Among various low-carbon industrial loads, low-carbon steelmaking systems based on direct reduced iron (DRI) and scrap-fed electric arc furnaces (EAFs) have emerged as a promising pathway for deep decarbonization in the iron and steel industry \cite{sun2020material}. However, the EAF steelmaking process still requires carbon injection to promote the carbon-oxygen reaction for foamy slag generation, thereby stabilizing arc heating and improving thermal efficiency \cite{liu2024current}. As a result, the DRI-EAF route still produces residual \ce{CO2} emissions ranging from 0.4-1.2 t \ce{CO2}/t crude steel, making it difficult to achieve zero-carbon steelmaking production \cite{wu2025technological}. Integrating methanol production into the DRI-EAF route (\ce{H2}-DRI-EAF-MeOH) provides a feasible pathway for circular carbon utilization and \ce{CO2} valorization. While this integration introduces intricate energy-material interactions that fundamentally reshape operational flexibility, existing studies \cite{lee2024critical,ranaboldo2024comprehensive, golmohamadi2022demand} rarely capture the deep synergies within such multi-energy coupled systems. Consequently, an explicit system-level formulation is imperative to accurately quantify the DR potential of the integrated \ce{H2}-DRI-EAF-MeOH system.\par
However, different process units in the integrated \ce{H2}-DRI-EAF-MeOH production system exhibit distinct operational characteristics. While the shaft furnace (SF) \cite{ji2026demand} and methanol synthesis reactor (MSR) \cite{yu2024optimal} can be reasonably described by quasi-steady-state models, existing optimization-oriented EAF models remain insufficient for capturing mass-energy coupling constraints and the operating feasible region. The EAF steelmaking process is influenced by multiple factors, including electrical energy input, charging quantities and temperatures of different feed materials, carbon injection, and flux addition \cite{liu2024current}. The interactions among these factors are constrained by strict thermodynamic, mass, and energy balances, leading to highly coupled and nonlinear operating characteristics. While simplified black-box models can reproduce the static input-output relationship between electricity consumption and steel output, they inherently overlook the coupled material-balance and process-state constraints that determine the practical EAF operating region \cite{su2023multi}. In particular, they cannot distinguish different electricity requirements associated with cold DRI and hot DRI charging from a pure-hydrogen SF. By contrast, detailed models based on mass and energy balances as well as reaction principles offer stronger physical interpretability, but their high nonlinearity makes them difficult to embed directly into optimization frameworks \cite{abadi2024review}. Therefore, it is essential to develop an EAF operating feasible region model that can represent physical constraints and energy-material balances while ensuring computational tractability for optimization-based scheduling and flexibility assessment. \par
Furthermore, existing DR evaluation frameworks are inadequate for assessing the DR potential of integrated \ce{H2}-DRI-EAF-MeOH systems. First, most studies focus on the flexibility of individual process units, such as alkaline electrolyzers (AE) \cite{bai2025enhancing}, EAFs \cite{zhang2016cost}, and refining furnaces \cite{wang2023quantifying}, making it difficult to evaluate the system-level DR potential of such a coupled production system. 
Second, existing DR studies in steel production systems \cite{matsveichuk2024models} often simplify equipment scheduling \cite{liu2025energy} and order fulfillment as static or penalty-based constraints \cite{gong2022integrated}, leaving order fulfillment effectively governed by unit capacity boundaries, which may lead to myopic DR scheduling decisions.
Third, existing industrial DR scheduling strategies still lack an explicit behavioral representation of operators’ inherent aversion to process deviations. Process dynamics \cite{esche2020dynamic}, safety limits \cite{bruns2021flexibility}, and operating feasibility \cite{otashu2020scheduling} are mainly considered, while operators’ subjective behavioral responses and asymmetric tolerance toward deviations from stable operating states are rarely explicitly characterized. Consequently, the disturbance cost and practical acceptability of key process-unit deviations remain insufficiently represented, which may lead to an overestimation of practically delivered DR capacity.
Therefore, a process-aware DR evaluation framework is needed to quantify system-level DR potential while ensuring order fulfillment and operational acceptability.\par
In summary, three critical research gaps remain in the assessment of DR potential in zero-carbon steelmaking systems: 1) at the system level, existing studies rarely represent the intricate multi-energy interactions, failing to accurately assess the DR potential of the integrated \ce{H2}-DRI-EAF-MeOH system; 2) at the modeling level, existing EAF models fail to simultaneously capture mass-energy coupling constraints and the operating feasible region; and 3) at the methodological level, existing DR evaluation frameworks lack process-aware scheduling formulations that account for dynamic order fulfillment and operators' inherent aversion to process deviations.\par
To address the aforementioned research gaps, this paper proposes a process-aware DR potential evaluation framework for \ce{H2}-DRI-EAF-MeOH. Specifically, the main contributions are as follows:\par
1) \textbf{System Level}: A closed-loop \ce{H2}-DRI-EAF-MeOH system is proposed to eliminate residual EAF carbon emissions via methanol synthesis. Furthermore, the integrated energy-material flows are explicitly formulated to capture the complex coupling interactions governing the system's DR potential. \par
2) \textbf{Modeling Level}: An optimization-compatible EAF operating feasible region model is developed based on mass- and energy-flow constraints. The model is calibrated and validated using actual production data from a pure-hydrogen DRI and EAF plant, yielding an average relative error of 4.1\%.\par
3) \textbf{Methodological Level}: A process-aware DR potential evaluation model is proposed for the integrated \ce{H2}-DRI-EAF-MeOH system. The proposed model incorporates a nonlinear asymmetric penalty formulation and an adaptive rolling mechanism, thereby explicitly characterizing operators' inherent aversion to process deviations while preventing myopic scheduling decisions.\par
4) \textbf{Assessment Level}: Dual-side DR evaluation metrics are formulated to comprehensively quantify the DR performance from both grid and load perspectives, thereby clarifying the grid-side effective delivered DR capacity and ramping risks while making the internal unit-level regulation behaviors explicitly observable and measurable. \par
The rest of this paper is organized as follows: Section II introduces the system architecture and component models. Section III presents the process-aware DR evaluation model and the dual-side evaluation metrics. Case studies are analyzed in Section IV. Finally, Section V concludes the paper.
\section{System Architecture and Component Modeling }
\subsection{System Architecture}
The architecture of the proposed \ce{H2}-DRI-EAF-MeOH system is illustrated in Fig. 1. The overall production process and material flows can be divided into four stages. First, the AE uses electricity to produce green hydrogen via water electrolysis. Second, the SF employs hot green hydrogen heated to 1050 °C as a reducing agent to remove oxygen from iron ore, producing either hot DRI (HDRI) or cold DRI (CDRI). Specifically, HDRI bypasses the waste heat boiler (WHB) to retain its sensible heat (detailed temperature specifications can be found in \cite{ji2025energy}), while CDRI is cooled through the WHB for heat recovery. Third, the EAF smelts CDRI, HDRI, and steel scrap into crude steel with carbon injection and flux addition. Meanwhile, the residual \ce{CO2} emissions are recovered by a carbon capture system and supplied as the carbon feedstock for the subsequent stage. Finally, the MSR converts a designated portion of the green hydrogen and the captured \ce{CO2} into MeOH, thereby enabling circular \ce{CO2} utilization within the integrated system. \par
Beyond the main material flows, the system also involves a multi-carrier energy flow network. Electricity serves as the dominant energy carrier and is supplied to the AE, EAF, and electric heating units to meet the electrical, thermal, and chemical energy demands of different production stages. The energy supply system comprises solar and wind power generation, a battery energy storage system (BESS), and the main grid, together with a three-stage expander for electricity recovery from high-pressure hydrogen expansion. The load side encompasses components such as AE, SF, EAF, MSR, low- and high-temperature electric heaters (LEH and HEH), and compressors. Furthermore, to mitigate spatio-temporal supply-demand mismatches among multiple coupled media, including electricity, hydrogen, heat, iron, and \ce{CO2}, various storage facilities are integrated into the system. These include low- and high-temperature thermal storage units (LTS and HTS), hydrogen storage tanks (HT), a CDRI silo (CDRIS), \ce{CO2} storage tanks (CST), and MeOH storage tanks (MST).
\begin{figure*}[!t]
    \centering
    \vspace{-2mm}
\includegraphics[width=0.98\textwidth,trim=0 4mm 0 0,clip]{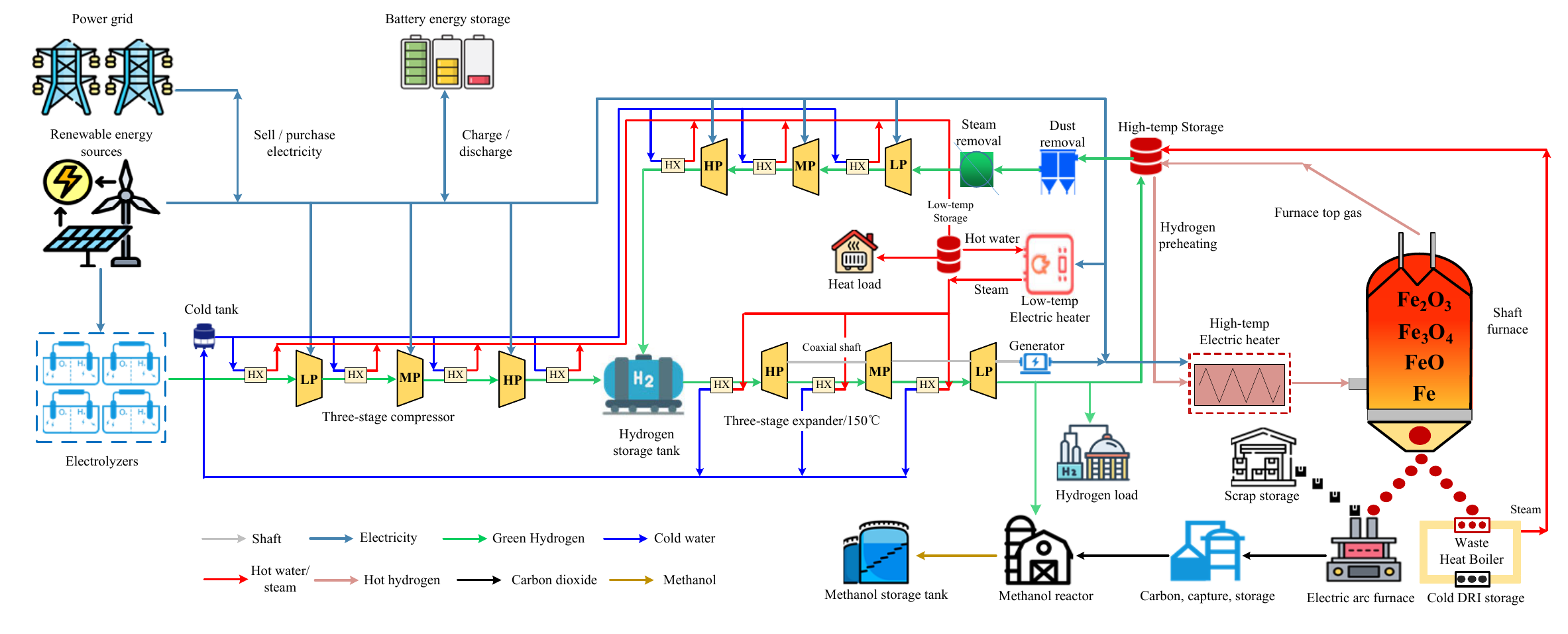}
    \vspace{-2mm}
    \caption{Hydrogen-integrated zero-carbon steel plant coupled with methanol production.}
    \vspace{-0.35cm}
\end{figure*}
\subsection{EAF Operating Feasible Region Modeling and Validation}\par
The EAF operating feasible region is formulated as a convex set under physical and mass-energy constraints, enabling its direct embedding into the optimization framework. To characterize this flexible region, the key operational state vector is defined as:
\begin{equation}
\boldsymbol{z}
=
\begin{bmatrix}
x_1 & x_2 & x_3 & P_{\mathrm{EAF}}
\end{bmatrix}^{\top},
\label{eq:eaf_state_vector}
\end{equation}
\begin{equation}
x_i = c_{\mathrm{p},i} M_i \Delta T_i,
\quad i \in \{1,2,3\},
\label{eq:sensible_heat}
\end{equation}
$x_i$, $c_{\mathrm{p},i}$, and $M_i$ denote the sensible heat, 
constant-pressure specific heat capacity, and mass of material $i$, respectively; 
$\Delta T_i$ is the temperature difference between the charging temperature 
of material $i$ and the reference temperature; and $P_{\mathrm{EAF}}$ denotes 
the electric power consumption of the EAF.\par
Because HDRI produced by the SF requires one full scheduling interval to complete production and delivery, the HDRI charged into the EAF at interval $t$ is determined by the SF output at interval $t-1$. This inter-temporal dependence is represented as a one-interval delay in the scheduling model. \par
The feasible operating region of the EAF can be expressed as
\begin{equation}
\boldsymbol{A}_{\mathrm{eq}}\boldsymbol{z}
=
\boldsymbol{b}_{\mathrm{eq}},
\label{eq:eaf_linear_constraints}
\end{equation}
\begin{equation}
\boldsymbol{A}_{\mathrm{eq}}
=
\begin{bmatrix}
\psi_{\mathrm{MT},1} & \psi_{\mathrm{MT},2} & \psi_{\mathrm{MT},3} & -1\\
\psi_{\mathrm{MI},1} & \psi_{\mathrm{MI},2} & \psi_{\mathrm{MI},3} & 0
\end{bmatrix},
\quad
\boldsymbol{b}_{\mathrm{eq}}
=
\begin{bmatrix}
0\\
M_{\mathrm{EAF}}
\end{bmatrix}.
\end{equation}
The two rows of $\boldsymbol{A}_{\mathrm{eq}}$ represent the EAF energy balance and metallic mass balance, respectively. The vector $\boldsymbol{b}_{\mathrm{eq}}$ is the corresponding right-hand side vector, and $M_{\mathrm{EAF}}$ denotes the target steel output per heat. \par
The feasible operating region of the EAF, together with the lower and upper bounds of the state variables, can be represented as the following convex polytope in H-representation:
\begin{equation}
\Gamma =
\{\boldsymbol{z}
\mid
\boldsymbol{A}_{\mathrm{eq}}\boldsymbol{z}
=
\boldsymbol{b}_{\mathrm{eq}},
\ 
\boldsymbol{z}_{\min}
\le
\boldsymbol{z}
\le
\boldsymbol{z}_{\max}
\}.
\end{equation}
\par In practical EAF operations, carbon powder and lime are injected to promote foamy slag formation and regulate slag basicity. The exothermic oxidation of injected carbon provides supplementary chemical heat, which directly affects the electricity demand $P_\mathrm{EAF}$. Since their dosages scale linearly with the target steel output under given steelmaking conditions, these auxiliary materials are treated as dependent process quantities rather than independent decision variables. Specifically, their thermal effects, such as the chemical heat released by carbon oxidation, are embedded in the constant coefficients of the energy balance matrix $\boldsymbol A_\mathrm{eq}$. The material consumption of these auxiliary inputs and the resulting residual \ce{CO2} emissions are quantified through the following auxiliary carbon-balance constraints at the material-flow level:
\begin{equation}
M_{\mathrm{EAF},\mathrm{CO}_2}^{t}
=
\psi_{\mathrm{C},c} M_{\mathrm{carbon}}^{t}
+
\psi_{\mathrm{C},l} M_{\mathrm{lime}}^{t}
+
\psi_{\mathrm{C},s} M_{\mathrm{scrap}}^{t},
\end{equation}
\begin{equation}
M_{\mathrm{carbon}}^{t}
=
\psi_{\mathrm{C},\mathrm{DRI}} M_{\mathrm{EAF}}^{t},
\end{equation}
\begin{equation}
M_{\mathrm{lime}}^{t}
=
\psi_{\mathrm{lime}} M_{\mathrm{EAF}}^{t}.
\end{equation}
$M_{\mathrm{EAF},\mathrm{CO}_2}^{t}$ denotes the \ce{CO2} emissions of the EAF at time interval $t$; 
$M_{\mathrm{scrap}}^{t}$, $M_{\mathrm{lime}}^{t}$, and $M_{\mathrm{carbon}}^{t}$ are the masses of steel scrap, lime, and carbon powder charged into the EAF at time interval $t$, respectively; $\psi$ denotes the corresponding emission, conversion, or consumption coefficients. \par
To address the highly coupled and nonlinear nature of EAF steelmaking, the proposed model translates the complex mass-energy balances into a convex polyhedral feasible region, as visualized in Fig. 2. Spanned by the charging quantities of hot DRI, cold DRI, and steel scrap, this 3D polytope explicitly reveals the substitution effect of cross-process hot material interactions on the operational flexibility of the EAF. Overall, the proposed formulation provides an optimization-tractable representation of the feasible operating boundary while preserving physical and process constraints, thereby supporting subsequent DR potential assessment of the integrated system. \par
To validate the proposed model, actual operating data from a commercial pure-hydrogen DRI production plant are used. 
The constant coefficients of the matrix $\boldsymbol{A}_{\mathrm{eq}}$ are extracted via a thermodynamic-based sensitivity analysis, incorporating specific material enthalpies, exothermic reaction heats, and slag formation energies derived from the actual chemical compositions of the DRI and steel scrap. The resulting matrix is given as follows:
\begin{equation}
\boldsymbol{A}_{\mathrm{eq}}=
\begin{bmatrix}
3.24\times10^{-4} &
3.11\times10^{-3} &
2.35\times10^{-3} &
-1 \\
9.79\times10^{-4} &
5.50\times10^{-3} &
5.26\times10^{-3} &
0
\end{bmatrix},
\end{equation}
The validation dataset contains the detailed chemical composition of the DRI product, including 92.31\% metallic Fe, 4.30\% FeO (corresponding to a metallization rate of 96.51\%), 2.28\% \ce{SiO2}, 0.68\% \ce{Al2O3}, 0.19\% \ce{MgO}, 0.09\% \ce{CaO}, and other minor elements. For the fully cold DRI charging base case, indicated by the black point in Fig. 2, the measured electricity consumption is 590 kWh/t crude steel. Under the same conditions, the boundary value predicted by the proposed model is 566 kWh/t crude steel. The resulting relative error of 4.1\% demonstrates that the proposed model can accurately estimate actual electricity demand while preserving physical consistency. 
\begin{figure}[b!] 
\centerline{\includegraphics[width=1 \columnwidth]{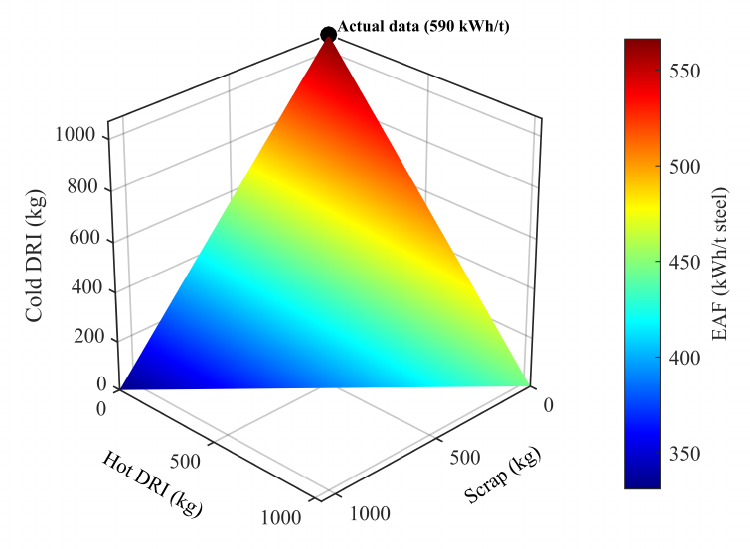}}
    \caption{EAF operating feasible region with electricity consumption and actual industrial data point.}
    \label{influence}
\end{figure}
\subsection{Modeling of Other Components}
\par \textbf{1) HEH.} To meet the high-temperature heat demand of the hydrogen-based SF for the endothermic reduction reaction, an HEH is employed to raise the hydrogen temperature to 1050 ℃. Accordingly, the electricity consumption of the HEH is determined by the corresponding high-temperature heat balance, which is given as follows:
\begin{subequations}
\begin{equation}
P_{\mathrm{HEH}}^{t}=
\psi_{\mathrm{Tth}} M_{\mathrm{DRI},\mathrm{p}}^{t}
-
\psi_{\mathrm{Th},\mathrm{re}}
( W_{\mathrm{ftg}}^{t} + W_{\mathrm{whb}}^{t})/ {
\psi_{\mathrm{Eh}}
},
\end{equation}
\begin{equation}
    W_{\mathrm{ftg}}^{t}=
\psi_{\mathrm{ftg}} M_{\mathrm{DRI},\mathrm{p}}^{t},
\end{equation}
\begin{equation}
    W_{\mathrm{whb}}^{t}=
\psi_{\mathrm{whb}} M_{\mathrm{DRI},\mathrm{Sin}}^{t}.
\end{equation}
\end{subequations}
$P_{\mathrm{HEH}}^{t}$ denotes the power of the HEH; 
$\psi_{\mathrm{Tth}}$ is the hydrogen heating demand coefficient for DRI production; $M_{\mathrm{DRI},\mathrm{p}}^{t}$ denotes the DRI output at time $t$; $W_{\mathrm{ftg}}^{t}$ denotes the recovered heat from furnace top gas; $\psi_{\mathrm{ftg}}$ is the unit heat recovery coefficient of furnace top gas; 
$\psi_{\mathrm{Th},\mathrm{re}}$ is the heat recovery efficiency; $W_{\mathrm{whb}}^{t}$ denotes the recovered heat from the WHB during HDRI cooling; 
$\psi_{\mathrm{whb}}$ and $\psi_{\mathrm{Eh}}$ represent the unit sensible heat recovery and electric heating conversion coefficients, respectively; and $M_{\mathrm{DRI},\mathrm{Sin}}^{t}$ denotes the storage inflow of CDRI at time $t$.
\par \textbf{2) SF.} Owing to the complex physicochemical reactions inside the hydrogen-based SF, its output response to load regulation exhibits pronounced dynamic lag characteristics. To represent the output transition during DRI discharge regulation, a first-order dynamic lag model is adopted following our prior work in \cite{ji2026demand}. The corresponding quasi-steady-state output, operating limits, and ramping constraints are expressed as follows:
\begin{subequations}
\begin{align}
M_{\mathrm{DRI},\mathrm{p}}^{t+1}
&=
\alpha_{\mathrm{DRI}} M_{\mathrm{QSS}}^{t}
+
\left( 1-\alpha_{\mathrm{DRI}} \right) M_{\mathrm{QSS}}^{t+1},
\label{eq:SF_dynamic_output}
\\
\alpha_{\mathrm{DRI}}
&=
e^{-\frac{\Delta t}{T_{\mathrm{DRI}}^{\mathrm{trans}}}},
\label{eq:SF_dynamic_alpha}
\\
M_{\mathrm{DRI},\mathrm{p}}^{k+1}
&=
M_{\mathrm{QSS}}^{k+1}
+
(
M_{\mathrm{QSS}}^{k}
-
M_{\mathrm{QSS}}^{k+1}
)
\alpha_{\mathrm{DRI}},
\label{eq:SF_dynamic_rolling}
\\
\psi_{\mathrm{MDRI}}^{\min} M_{\mathrm{DRI},\mathrm{dis}}^{\max}
&\le
M_{\mathrm{DRI},\mathrm{QSS}}^{t}
\le
\psi_{\mathrm{MDRI}}^{\max} M_{\mathrm{DRI},\mathrm{dis}}^{\max},
\label{eq:SF_qss_bounds}
\\
\psi_{\mathrm{down}}^{\mathrm{ramp}} M_{\mathrm{DRI},\mathrm{dis}}^{\max}
&\le
M_{\mathrm{QSS}}^{k+1}
-
M_{\mathrm{QSS}}^{k}
\le
\psi_{\mathrm{up}}^{\mathrm{ramp}} M_{\mathrm{DRI},\mathrm{dis}}^{\max}.
\label{eq:SF_ramping_limits}
\end{align}
\end{subequations}
$M_{\mathrm{QSS}}^{t}$ denotes the quasi-steady-state DRI discharge of the SF at time $t$; $T_{\mathrm{DRI}}^{\mathrm{trans}}$ is the transition time constant of the SF; $\psi_{\mathrm{MDRI}}^{\min}$ and $\psi_{\mathrm{MDRI}}^{\max}$ are the minimum and maximum production coefficients, respectively; 
$\psi_{\mathrm{down}}^{\mathrm{ramp}}$ and $\psi_{\mathrm{up}}^{\mathrm{ramp}}$ denote the downward and upward ramping coefficients, respectively; and $M_{\mathrm{DRI},\mathrm{dis}}^{\max}$ is the maximum discharge of the SF. \par
Driven by EAF mass demand and storage scheduling, the DRI produced by the SF is routed to either the hot-charging stream or the cold silo. The cold silo must satisfy the following inter-temporal mass-balance constraints:
\begin{subequations}\label{eq:DRI_storage}
\begin{equation}
M_{\mathrm{DRI},\mathrm{p}}^{t}
=
M_{\mathrm{DRI},\mathrm{hot,out}}^{t}
+
M_{\mathrm{DRI},\mathrm{Sin}}^{t},
\end{equation}
\begin{equation}
M_{\mathrm{DRI},\mathrm{EAF,hot}}^{t}
=
M_{\mathrm{DRI},\mathrm{hot,out}}^{t-1},
\end{equation}
\begin{equation}
S_{\mathrm{DRI}}^{t+1}
=
S_{\mathrm{DRI}}^{t}
+
M_{\mathrm{DRI},\mathrm{Sin}}^{t}
-
M_{\mathrm{DRI},\mathrm{EAF,cold}}^{t},
\end{equation}
\begin{equation}
S_{\mathrm{DRI}}^{1}
=
S_{\mathrm{DRI}}^{T+1}
=
0.5 S_{\mathrm{DRI}}^{\max}.
\end{equation}
\end{subequations}
$M_{\mathrm{DRI},\mathrm{hot,out}}^{t}$ denotes the mass of hot-charged DRI produced by the SF at time $t$; $M_{\mathrm{DRI},\mathrm{EAF,hot}}^{t}$ is the mass of HDRI charged into the EAF at time $t+1$; 
$M_{\mathrm{DRI},\mathrm{EAF,cold}}^{t-1}$ is the mass of CDRI required by the EAF at time $t$; 
and $S_{\mathrm{DRI}}^{t}$ denotes the CDRI storage state at the end of time period $t$.
\par \textbf{3) MSR.} Similar to the hydrogen-based SF, the MSR exhibits first-order dynamic lag during load regulation. Its output-transition and ramping constraints follow the same mathematical structure as Eqs. (11a)-(11e) and are therefore omitted here for brevity, while the detailed formulations can be found in \cite{yu2025novel}. Additionally, based on reaction stoichiometry, the hydrogen consumption and captured \ce{CO2} demand for MeOH synthesis are formulated as follows:
\begin{subequations}
\begin{align}
M_{\mathrm{Hmetha}}^{t}
&=
\psi_{\mathrm{Hmetha}} M_{\mathrm{metha},\mathrm{p}}^{t},
\\
M_{\mathrm{Cmetha}}^{t}
&=
\psi_{\mathrm{Cmetha}} M_{\mathrm{metha},\mathrm{p}}^{t}.
\end{align}
\end{subequations}
$M_{\mathrm{metha},\mathrm{p}}^{t}$ denotes the MeOH output of the MSR at time $t$; 
$M_{\mathrm{Hmetha}}^{t}$ and $M_{\mathrm{Cmetha}}^{t}$ are the hydrogen demand and \ce{CO2} demand for MeOH synthesis at time $t$, respectively; 
and $\psi_{\mathrm{Hmetha}}$ and $\psi_{\mathrm{Cmetha}}$ are the corresponding hydrogen and \ce{CO2} consumption coefficients.
\par \textbf{4) Generalized energy storage model}. To mitigate the operational uncertainties induced by RES intermittency and electricity price fluctuations, the proposed system integrates multiple storage units, including the BESS, LTS, HT, CDRIS, ScS, and CST. These heterogeneous storage units are modeled within a unified generalized storage framework:
\begin{subequations}
\begin{equation}
E_{m}^{t+1}
=
E_{m}^{t}
+
(
\psi_{m}^{\mathrm{ch}} P_{m,\mathrm{ch}}^{t}
-
\frac{P_{m,\mathrm{dis}}^{t}}{\psi_{m}^{\mathrm{dis}}}
)\Delta t,
\end{equation}
\begin{equation}
E_{m}^{\min}
\le
E_{m}^{t}
\le
E_{m}^{\max}
\end{equation}
\begin{equation}
0
\le
P_{m,\mathrm{ch}}^{t}
\le
b_{m,\mathrm{ch}}^{t} P_{m,\mathrm{ch}}^{\max},
\end{equation}
\begin{equation}
0
\le
P_{m,\mathrm{dis}}^{t}
\le
b_{m,\mathrm{dis}}^{t} P_{m,\mathrm{dis}}^{\max},
\end{equation}
\begin{equation}
b_{m,\mathrm{ch}}^{t}
+
b_{m,\mathrm{dis}}^{t}
\le
1.
\end{equation}
\end{subequations}
$m \in \{\mathrm{BESS}, \mathrm{LTS}, \mathrm{HT}, \mathrm{CDRIS}, \mathrm{ScS}, \mathrm{CST}\}$ indexes the storage types; 
$E_{m}^{t}$ is the state of storage (SoC) at time $t$; 
$P_{m,\mathrm{ch}}^{t}$ and $P_{m,\mathrm{dis}}^{t}$ denote the charging and discharging rates, respectively; $\psi_{m}^{\mathrm{ch}}$ and $\psi_{m}^{\mathrm{dis}}$ are the charging and discharging efficiencies, respectively; and $b_{m,\mathrm{ch}}^{t}$ and $b_{m,\mathrm{dis}}^{t}$ are mutually exclusive binary state variables used to prevent simultaneous charging and discharging within the same time period. \par
For brevity, the detailed models of the AE, compressor, and expander are not repeated here and can be found in \cite{yu2024optimal, ji2025energy}.
\section{Process-aware DR potential evaluation model and dual-side assessment metrics}
The DR potential of the \ce{H2}-DRI-EAF-MeOH system is evaluated by comparing the baseline trajectory with a DR-oriented schedule obtained through day-ahead offering and intra-day implementation.
\subsection{Baseline scheduling}
In the day-ahead baseline (BD) dispatch stage, forecast information is used to determine the baseline production schedule of the proposed system. To reflect the inherent requirements for process stability and production continuity commonly observed in industrial practice, the key production units are maintained at stable operating states,  The mathematical formulation is given as follows:
\begin{subequations}
\begin{equation}
\min F_{\mathrm{ec}}^{\mathrm{BD}},
\end{equation}
\begin{equation}
F_{\mathrm{ec}}^{\mathrm{BD}}
=
\sum_{t\in\mathcal{T}}
\left(
C_{\mathrm{grid}}^{t}
+
C_{\mathrm{op}}^{t}
-
R_{\mathrm{sell}}^{t}
\right)
+
\rho_{\mathrm{peak}} P_{\mathrm{buy},\mathrm{peak}},
\end{equation}
\begin{equation}
C_{\mathrm{grid}}^{t}
=
\rho_{\mathrm{buy}}^{t}P_{\mathrm{buy}}^{t}
+
\rho_{\mathrm{curt}}P_{\mathrm{curt}}^{t},
\quad
\forall t\in\mathcal{T}
\end{equation}
\begin{equation}
C_{\mathrm{op}}^{t}
=
\sum_{i\in\Omega_{\mathrm{E}}}
C_{i}^{\mathrm{E}}P_{i}^{t}
+
\sum_{j\in\Omega_{\mathrm{M}}}
C_{j}^{\mathrm{M}}M_{j}^{t},
\quad
\forall t\in\mathcal{T}
\end{equation}
\begin{equation}
R_{\mathrm{sell}}^{t}
=
\rho_{\mathrm{sell}}^{t}P_{\mathrm{sell}}^{t}
+
\rho_{\mathrm{thl}}\mathrm{Thl}_{\mathrm{load}}^{t}
+
\rho_{\mathrm{hl}}\mathrm{Hl}_{\mathrm{sell}}^{t},
\quad
\forall t\in\mathcal{T}
\end{equation}
\begin{equation}
P_{\mathrm{buy},\mathrm{peak}}
\ge
P_{\mathrm{buy}}^{t},
\quad
\psi_{u}^{t}
=
\psi_{u}^{\mathrm{base}},
\quad
\forall u \in U_{\mathrm{core}},\ \forall t\in\mathcal{T}
\end{equation}
\begin{equation}
\sum_{t\in\mathcal{T}} P_{\mathrm{sell}}^{t}
\le
\psi_{\mathrm{sell}}
\sum_{t\in\mathcal{T}}
\left(
P_{\mathrm{s}}^{t}
+
P_{\mathrm{w}}^{t}
\right),
\end{equation}
\begin{equation}
\sum_{t\in\mathcal{T}} M_{v}^{t}
=
M_{v}^{\mathrm{order}},
\qquad
\forall v\in\{\mathrm{SF},\mathrm{EAF}\}
\end{equation}
\begin{equation}
\mathrm{s.t.}\quad (1)\text{--}(14),\ (17).
\end{equation}
\end{subequations}\par
The intra-day baseline (BI) scheduling is implemented using a rolling MPC strategy, in which dispatch decisions are updated with the latest forecasts while key production units track their day-ahead scheduled stable states. Following the methodology in \cite{huang2025grid},   
the SoC references are recursively updated using realized conditions and historical perfect-information trajectories, providing long-term guidance for intra-day optimization and mitigating myopic decisions. 
The formulation is given as follows:
\begin{subequations}
\begin{equation}
\min F_{\mathrm{ec}}^{k,\mathrm{BI}},
\end{equation}
\begin{equation}
\begin{aligned}
F_{\mathrm{ec}}^{k,\mathrm{BI}}
&=
\sum_{t \in \mathcal{T}_{k}^{L}}
(
C_{\mathrm{grid}}^{t,k}
+
C_{\mathrm{op}}^{t,k}
-
R_{\mathrm{sell}}^{t,k}
\\
&+
\lambda_{\mathrm{rf}}
\left\lvert
\mathrm{SoC}_{n}^{t}
-
\mathrm{SoC}_{n,\mathrm{rf}}^{t}
\right\rvert)
+
\rho_{\mathrm{peak}} P_{\mathrm{buy},\mathrm{peak}},
\end{aligned}
\end{equation}
\begin{equation}
\mathrm{SoC}_{n,\mathrm{rf}}^{t}
=
\sum_{s=1}^{S}
\rho^{t,s}
\,\mathrm{SoC}_{n}^{t,s},
\quad
\sum_{s=1}^{S}\rho^{t,s}
=
1,
\end{equation}
\begin{equation}
\rho^{t,s}
=
\frac{K_t^{s}}
{\sum_{s'=1}^{S}K_t^{s'}},
\quad
K_t^{s}
=
\prod_{q\in\mathcal{Q}}
K_t^{q,s},
\end{equation}
\begin{equation}
K_t^{q,s}
=
\exp\!(
-\frac{
\left\lVert
\boldsymbol{g}_{q,[t]}
-
\boldsymbol{g}_{q,[t]}^{\,s}
\right\rVert_2^2
}{
2\sigma_q^2(t+1)
}
),
\end{equation}
\begin{equation}
C_{\mathrm{grid}}^{t,k}
=
\rho_{\mathrm{buy}}^{t,k}P_{\mathrm{buy}}^{t,k}
+
\rho_{\mathrm{curt}}P_{\mathrm{curt}}^{t,k},
\end{equation}
\begin{equation}
C_{\mathrm{op}}^{t,k}
=
\sum_{i\in\Omega_{\mathrm{E}}}
C_i^{\mathrm{E}}P_i^{t,k}
+
\sum_{j\in\Omega_{\mathrm{M}}}
C_j^{\mathrm{M}}M_j^{t,k},
\end{equation}
\begin{equation}
R_{\mathrm{sell}}^{t,k}
=
\rho_{\mathrm{sell}}^{t,k}P_{\mathrm{sell}}^{t,k}
+
\rho_{\mathrm{thl}}\mathrm{Thl}_{\mathrm{load}}^{t,k}
+
\rho_{\mathrm{hl}}\mathrm{Hl}_{\mathrm{sell}}^{t,k},
\end{equation}
\begin{equation}
P_{\mathrm{buy},\mathrm{peak}}
\ge
P_{\mathrm{buy}}^{\tau,\mathrm{real}},
\qquad
\forall \tau = 1,\ldots,k-1
\end{equation}
\begin{equation}
X_m^{k,\mathrm{ID}}
=
X_m^{k,\mathrm{real}},
\qquad
\forall m \in \mathcal{X}_{\mathrm{state}}
\end{equation}
\begin{equation}
\psi_u^{t,k,\mathrm{BI}}
=
\psi_u^{\mathrm{base}},
\qquad
\forall u \in U_{\mathrm{core}},\ \forall t \in \mathcal{T}_{k}^{L}
\end{equation}
\begin{equation}
P_{\mathrm{sell},\max}^{k}
=
\psi_{\mathrm{sell}}(
\sum_{\tau=1}^{k-1}
(
P_{\mathrm{s}}^{\tau,\mathrm{real}}
+
P_{\mathrm{w}}^{\tau,\mathrm{real}}
)+
\sum_{t \in \mathcal{T}_{k}^{L}}
\left(
P_{\mathrm{s}}^{t,k}
+
P_{\mathrm{w}}^{t,k}
\right)),
\end{equation}
\begin{equation}
P_{\mathrm{sell},\mathrm{cum}}(k)
=
\sum_{\tau=1}^{k-1}
P_{\mathrm{sell}}^{\tau,\mathrm{real}},
\end{equation}
\begin{equation}
P_{\mathrm{sell},\mathrm{rem}}(k)
=
P_{\mathrm{sell},\max}^{k}
-
P_{\mathrm{sell},\mathrm{cum}}(k),
\end{equation}
\begin{equation}
\sum_{t \in \mathcal{T}_{k}^{L}}
P_{\mathrm{sell}}^{t,k,\mathrm{ID}}
\le
\max
\left(
0,\,
P_{\mathrm{sell},\mathrm{rem}}(k)
\right),
\end{equation}
\begin{equation}
\text{s.t.}\quad (1)\text{--}(14),\ (15g),\ (17).
\end{equation}
\end{subequations}
$F_{\mathrm{ec}}$ denotes the economic objective.
$C_{\mathrm{grid}}^{t}$, $C_{\mathrm{op}}^{t}$, and $R_{\mathrm{sell}}^{t}$ denote grid electricity cost, operating cost, and sales revenue, respectively. 
$\rho$ and $C$ are the corresponding price and unit operating-cost coefficients. 
$P_i^t$ and $M_j^t$ denote the electricity consumption of unit $i\in\Omega_{\mathrm{E}}$ and the material flow of unit $j\in\Omega_{\mathrm{M}}$, respectively. 
$\mathrm{Thl}_{\mathrm{load}}^{t}$ denotes the thermal load, and $P_{\mathrm{buy},\mathrm{peak}}$ denotes the peak grid-purchased power for the capacity charge. 
$\psi_u^{\mathrm{base}}$ denotes the baseline operating state of key unit $u$, and $M_v^{\mathrm{order}}$ denotes the production order of unit $v$. 
$U_{\mathrm{core}}=\{\mathrm{AE},\mathrm{SF},\mathrm{EAF},\mathrm{MSR}\}$ denotes the set of key process units. 
$\mathrm{SoC}_{n,\mathrm{rf}}^{t}$, $\rho^{t,s}$, $K_t^{s}$, and $K_t^{q,s}$ denote the state reference, scenario weight, composite kernel similarity, and feature-wise kernel similarity, respectively. 
$\mathcal{Q}=\{\mathrm{w},\mathrm{s},\rho_{\mathrm{buy}},\rho_{\mathrm{sell}},\mathrm{Hl},\mathrm{Thl}\}$. 
$P_{\mathrm{sell},\max}^{k}$, $P_{\mathrm{sell},\mathrm{cum}}(k)$, and $P_{\mathrm{sell},\mathrm{rem}}(k)$ denote the maximum allowable electricity sales, cumulative realized sales, and remaining sales quota at rolling instant $k$, respectively. $\psi_{\mathrm{sell}}$ denotes the on-site RES consumption ratio. \par
In addition, the proposed system is subject to constraints on electricity trading, on-site RES utilization, and real-time power balance, as follows:
\begin{subequations}
\begin{equation}
0
\le
P_{\mathrm{sell}}^{t}
\le
b_{\mathrm{grid}}^{t}M,
\end{equation}
\begin{equation}
0
\le
P_{\mathrm{buy}}^{t}
\le
\left(1-b_{\mathrm{grid}}^{t}\right)M,
\end{equation}
\begin{equation}
P_{\mathrm{supply}}^{t}
=
P_{\mathrm{demand}}^{t},
\end{equation}
\begin{equation}
P_{\mathrm{supply}}^{t}
=
P_{\mathrm{s}}^{t}
+
P_{\mathrm{w}}^{t}
+
P_{\mathrm{buy}}^{t}
+
P_{\mathrm{exp}}^{t}
+
P_{\mathrm{BESS},\mathrm{dis}}^{t},
\end{equation}
\begin{equation}
\begin{aligned}
P_{\mathrm{demand}}^{t}
&=
P_{\mathrm{sell}}^{t}
+
P_{\mathrm{EAF}}^{t}
+
P_{\mathrm{BESS},\mathrm{ch}}^{t}
+
P_{\mathrm{AE}}^{t}
\\
&\quad
+
P_{\mathrm{comp}}^{t}
+
P_{\mathrm{HEH}}^{t}
+
P_{\mathrm{LEH}}^{t}
+
P_{\mathrm{CCS}}^{t}
+
P_{\mathrm{curt}}^{t},
\end{aligned}
\end{equation}
\begin{equation}
P_{\mathrm{comp}}^{t}
=
P_{\mathrm{AE}}^{t}\psi_{\mathrm{ERcomp}}
+
M_{\mathrm{DRI},\mathrm{p}}^{t}\psi_{\mathrm{ECcomp}},
\end{equation}
\begin{equation}
P_{\mathrm{exp}}^{t}
=
\left(
M_{\mathrm{DRI},\mathrm{p}}^{t}\psi_{\mathrm{H},\mathrm{DRI}}
+
Hl_{\mathrm{sell}}^{t}
\right)
\psi_{\mathrm{Eexp}}.
\end{equation}
\end{subequations}
$b_{\mathrm{grid}}^{t}$ is a binary variable and $M$ is a large positive constant. 
The compressor load and expander output are determined by the corresponding conversion coefficients: $\psi_{\mathrm{ERcomp}}$ and $\psi_{\mathrm{ECcomp}}$ denote the compression electricity coefficients for electrolytic hydrogen and recovered hydrogen, respectively; $\psi_{\mathrm{H},\mathrm{DRI}}$ denotes the hydrogen demand coefficient for DRI production; and $\psi_{\mathrm{Eexp}}$ denotes the electricity generation coefficient of the three-stage expander.
\subsection{Process-aware DR potential scheduling} In the day-ahead DR (DD) dispatch stage, the baseline schedule is used as the reference for evaluating process deviations of key units. A nonlinear asymmetric process-deviation penalty model is developed, in which deviations are normalized by operational limits to capture operators' different tolerance levels for upward and downward adjustments. After obtaining the DD schedule, the deviation between the baseline and DR-oriented net-load trajectories is converted into the magnitude and direction of the day-ahead DR commitment, as follows:
\begin{subequations}
\begin{equation}
\min F_{\mathrm{ec}}^{\mathrm{DD}}
+
\lambda_{\mathrm{p}}D_{\mathrm{p}}^{\mathrm{DD}},
\end{equation}
\begin{equation}
D_{\mathrm{p}}^{\mathrm{DD}}
=
\sum_{t\in\mathcal{T}}
\sum_{u\in U_{\mathrm{core}}}
\omega_u
\sum_{s\in\{+,-\}}
\alpha_{u,s}
\Phi_{u,s}^{t,\mathrm{DD}},
\end{equation}
\begin{equation}
\Phi_{u,s}^{t,\mathrm{DD}}
=
\exp\!\left(
\beta_{u,s}
\big(
\sigma_s \Delta \psi_u^{t,\mathrm{DD}}
-
\epsilon_{u,s}
\big)_+
\right)
-
1,
\end{equation}
\begin{equation}
\Delta \psi_u^{t,\mathrm{DD}}
=
\frac{
\psi_u^{t,\mathrm{DD}}
-
\psi_u^{t,\mathrm{BD}}
}{
\psi_{u,\max}
},
\end{equation}
\begin{equation}
\sigma_{+}=1,
\quad
\sigma_{-}=-1,
\quad
(x)_+=\max(0,x),
\end{equation}
\begin{equation}
B^t
=
P_{\mathrm{netload}}^{t,\mathrm{BS}}
-
P_{\mathrm{netload}}^{t,\mathrm{DR}},
\qquad
P_{\mathrm{mag}}^{t,\mathrm{DA}}
=
\left|B^t\right|,
\end{equation}
\begin{equation}
P_{\mathrm{netload}}=P_{\mathrm{buy}}-P_{\mathrm{sell}},
\end{equation}
\begin{equation}
d^t
=
\mathrm{sign}(B^t),
\qquad
P_{\mathrm{offer}}^{t,\mathrm{DA}}
=
d^t P_{\mathrm{mag}}^{t,\mathrm{DA}},
\end{equation}
\begin{equation}
\mathrm{s.t.}\quad
(1)\text{--}(14),\ 
(15c)\text{--}(15h),\ 
(17).
\end{equation}
\end{subequations}\par
The intra-day DR (DI) scheduling is formulated as a rolling MPC strategy. 
Its objective includes the economic cost, state-reference tracking term, offer-shortfall penalty, and process-deviation penalty relative to the BI benchmark.\par 
Additionally, an adaptive rolling mechanism (ARM) is proposed in (19e)--(19f) based on proportional residual tracking. 
By updating the remaining production requirement from realized cumulative output and reallocating it over the rolling horizon, the ARM prevents the continuous deferral of production tasks in pursuit of short-term cost savings, thereby mitigating MPC myopia under RES and electricity-price uncertainties.
\begin{subequations}
\begin{equation}
\min \;
F_{\mathrm{ec}}^{k,\mathrm{DI}}
+
\lambda_{\mathrm{rf}}D_{\mathrm{rf}}^{k,\mathrm{DI}}
+
\lambda_{\mathrm{s}}D_{\mathrm{s}}^{k,\mathrm{DI}}
+
\lambda_{\mathrm{p}}D_{\mathrm{p}}^{k,\mathrm{DI}},
\end{equation}
\begin{equation}
D_{\mathrm{p}}^{k,\mathrm{DI}}
=
\sum_{t\in\mathcal{T}_k^L}
\sum_{u\in U_{\mathrm{core}}}
\omega_u
\sum_{s\in\{+,-\}}
\alpha_{u,s}
\Phi_{u,s}^{t,k,\mathrm{DI}},
\end{equation}
\begin{equation}
D_{\mathrm{s}}^{k,\mathrm{DI}}
=
\sum_{t\in\mathcal{T}_k^{L}}
\max\!\left(
0,\;
P_{\mathrm{mag}}^{t,\mathrm{DA}}
-
d^{t} B^{t,k,\mathrm{DI}}
\right),
\end{equation}
\begin{equation}
B^{t,k,\mathrm{DI}}
=
P_{\mathrm{netload}}^{t,k,\mathrm{BI}}
-
P_{\mathrm{netload}}^{t,k,\mathrm{DI}},
\end{equation}
\begin{equation}
\sum_{t \in \mathcal{T}_{k}^{L}}
M_{v}^{t,k,\mathrm{DI}}
\le
M_{v}^{\mathrm{order}}
-
\sum_{\tau=1}^{k-1}
M_{v}^{\tau,\mathrm{real}},
\end{equation}
\begin{equation}
\sum_{t \in \mathcal{T}_{k}^{L}}
M_{v}^{t,k,\mathrm{DI}}
\ge
\frac{L}{T-k+1}
\left(
M_{v}^{\mathrm{order}}
-
\sum_{\tau=1}^{k-1}
M_{v}^{\tau,\mathrm{real}}
\right),
\end{equation}
\begin{equation}
F_{\mathrm{global}}^{\mathrm{DI}}
=
F_{\mathrm{ec,global}}^{\mathrm{DI}}
+
\lambda_{\mathrm{s}}D_{\mathrm{s}}^{\mathrm{global}}
+
\lambda_{\mathrm{p}}D_{\mathrm{p}}^{\mathrm{global}}
+
C_{\mathrm{H}_2,\mathrm{bf}},
\end{equation}
\begin{equation}
C_{\mathrm{H}_2,\mathrm{bf}}
=
\rho_{\mathrm{H}_2,\mathrm{bf}}
\left(
S_{\mathrm{HT}}^{T+1,\mathrm{BI}}
-
S_{\mathrm{HT}}^{T+1,\mathrm{DI}}
\right),
\end{equation}
\begin{equation}
\mathrm{s.t.}\quad
(1)\text{--}(14),\ 
(15c)\text{--}(15h),\ (16l)\text{--}(16o),\
(17).
\end{equation}
\end{subequations}
$F_{\mathrm{ec}}^{\mathrm{DD}}$, $F_{\mathrm{ec}}^{k,\mathrm{DI}}$, $F_{\mathrm{global}}^{\mathrm{DI}}$, and $F_{\mathrm{ec,global}}^{\mathrm{DI}}$ denote the DD economic objective, DI rolling economic objective, realized full-horizon objective, and corrected realized full-horizon economic cost, respectively. 
$\lambda_i$ is the penalty coefficient of term $i$; $D_{\mathrm{rf}}^{k,\mathrm{DI}}$, $D_{\mathrm{s}}^{k,\mathrm{DI}}$, and $D_{\mathrm{p}}^{k,\mathrm{DI}}$ denote the reference-tracking, offer-shortfall, and process-deviation penalty, respectively. 
$\Delta\psi_u$, $\Phi_{u,s}$, $\omega_u$, $\alpha_{u,s}$, $\beta_{u,s}$, and $\epsilon_{u,s}$ denote the normalized deviation, nonlinear penalty, unit weight, penalty amplitude, steepness, and tolerance threshold, respectively. 
$\psi_u$ and $\psi_{u,\max}$ denote the operating variable and rated normalization capacity of key unit $u$, respectively. $L$ is the rolling horizon length, $T$ is the full scheduling horizon length, $k$ is the current rolling instant, and $\tau$ is the realized time-period index. 
$P_{\mathrm{mag}}^{t,\mathrm{DA}}$, $d^t$, and $P_{\mathrm{offer}}^{t,\mathrm{DA}}$ denote the magnitude, direction, and signed value of the day-ahead DR offer, respectively; $B^t$ and $B^{t,k,\mathrm{DI}}$ denote the day-ahead offered and intra-day delivered DR capacities, respectively. 
$M_v^{t,k,\mathrm{DI}}$ and $M_v^{\tau,\mathrm{real}}$ denote planned and realized production, respectively; $C_{\mathrm{H}_2,\mathrm{bf}}$ and $\rho_{\mathrm{H}_2,\mathrm{bf}}$ denote the terminal hydrogen backfilling cost and unit production cost.
\vspace{-0.5cm}
\subsection{Dual-side DR potential evaluation metrics}
\par
1) \textbf{Average effective delivered DR capacity}: This metric measures the average DR capacity delivered in the day-ahead offered direction, reflecting the executable regulation capability of the proposed framework.
\begin{subequations}
\begin{equation}
\Omega_{\mathrm{offer}}
=
\left\{
t\in\mathcal{T}
\mid
B_{\mathrm{offer}}^{t}>0
\right\},
\end{equation}
\begin{equation}
B_{\mathrm{eff}}^{t}
=
\min\{
B_{\mathrm{offer}}^{t},
\ 
\max(
0,\ 
d^{t}
(
P_{\mathrm{netload}}^{t,\mathrm{BI}}
-
P_{\mathrm{netload}}^{t,\mathrm{DI}}
)
)
\},
\end{equation}
\begin{equation}
\bar{B}_{\mathrm{eff}}
=
\frac{1}{\left|\Omega_{\mathrm{offer}}\right|}
\sum_{t\in\Omega_{\mathrm{offer}}}
B_{\mathrm{eff}}^{t}.
\end{equation}
\end{subequations}
\par 2) \textbf{Recovery ramping rate}: This metric measures the largest upward change in grid power caused by post-DR load rebound.
\begin{equation}
RR^{+}
=
\max_{t\in\mathcal{T}^{+}}
\left(
P_{\mathrm{netload}}^{t+1,\mathrm{DI}}
-
P_{\mathrm{netload}}^{t,\mathrm{DI}}
\right).
\end{equation}
\par 3) \textbf{Dynamic RES-load matching degree}: This metric evaluates how closely internal load variations follow RES fluctuations, thereby reflecting the system's internal RES-load matching capability.
\begin{subequations}
\begin{equation}
\eta_{\mathrm{match}}
=
1
-
\frac{
\sum_{t\in\mathcal{T}^{+}}
\left|
\Delta P_{\mathrm{load}}^{t}
-
\Delta P_{\mathrm{RE}}^{t}
\right|
}{
\sum_{t\in\mathcal{T}^{+}}
\left|
\Delta P_{\mathrm{load}}^{t}
\right|
+
\sum_{t\in\mathcal{T}^{+}}
\left|
\Delta P_{\mathrm{RE}}^{t}
\right|
},
\end{equation}
\begin{equation}
\begin{aligned}
P_{\mathrm{load}}^{t}
=
P_{\mathrm{EAF}}^{t}
+
P_{\mathrm{BESS},\mathrm{ch}}^{t}
+
P_{\mathrm{AE}}^{t}
+
P_{\mathrm{comp}}^{t}
&\\
+
P_{\mathrm{HEH}}^{t}
+
P_{\mathrm{LEH}}^{t}
+
P_{\mathrm{CCS}}^{t},
\end{aligned}
\end{equation}
\begin{equation}
P_{\mathrm{RE}}^{t}
=
P_{\mathrm{s}}^{t}
+
P_{\mathrm{w}}^{t},
\end{equation}
\begin{equation}
\Delta P_{\mathrm{load}}^{t}
=
P_{\mathrm{load}}^{t}
-
P_{\mathrm{load}}^{t-1},
\quad
\forall t\in\mathcal{T}^{+},
\end{equation}
\begin{equation}
\Delta P_{\mathrm{RE}}^{t}
=
P_{\mathrm{RE}}^{t}
-
P_{\mathrm{RE}}^{t-1},
\quad
\forall t\in\mathcal{T}^{+}.
\end{equation}
\end{subequations}
$\mathcal{T}^{+}=\{1,2,\ldots,T-1\}$.
$P_{\mathrm{load}}^t$ is the actual active power load of the internal units. Mathematically, according to the absolute value inequality $|\Delta P_{\mathrm{load}}^t - \Delta P_{\mathrm{RE}}^t| \le |\Delta P_{\mathrm{load}}^t| + |\Delta P_{\mathrm{RE}}^t|$, the value of the fractional term is always constrained within $[0, 1]$. Consequently, the resulting matching degree $\eta_{\mathrm{match}}$ is strictly bounded between 0 and 1.\par
4) \textbf{Normalized regulation intensity (NRI)}: This metric characterizes the cumulative regulation intensity of each flexible unit by normalizing its step-to-step fluctuations, making its adjustment actions observable and measurable.
\begin{equation}
\mathrm{NRI}_{u}
=
\frac{1}{\psi_{u,\max}}
\sum_{t\in\mathcal{T}^{+}}
\left|
\psi_{u}^{t+1}
-
\psi_{u}^{t}
\right|,
\quad
\forall u\in U_{\mathrm{core}}.
\end{equation}
\section{Case Study}\label{case study} 
\subsection{Setups}
The profiles of RES generation and real-time electricity prices are shown in Fig. 3. These time-series inputs are treated as exogenous representative profiles for DR potential assessment: the day-ahead stage incorporates a 10\% prediction error, while the intra-day MPC stage uses an 8-step look-ahead horizon with a 5\% prediction error. The proposed model is implemented in Python and solved using Gurobi 12.0.3 on a computer with an AMD Ryzen 7 7840S processor.
\begin{figure}[htbp]
    \centering
    \vspace{-2mm}
    \includegraphics[width=1.0\columnwidth,trim=0 3mm 0 0,clip]{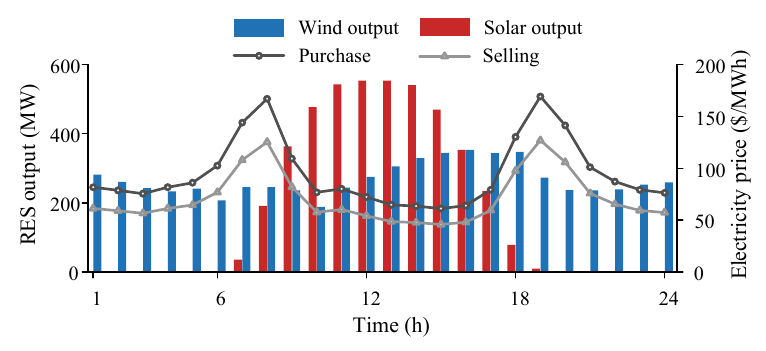}
    \vspace{-2mm}
    \caption{RES output and electricity-price profiles.}
    \label{fig:res_price}
    \vspace{-0.35cm}
\end{figure}
\vspace{-0.75cm}
\subsection{Performance comparison of baseline and DR schemes}
Compared with the BI, the proposed \ce{H2}-DRI-EAF-MeOH system achieves a lower operating cost and unlocks significant DR potential under the DI scenario with $\lambda_{\mathrm{p}}=50 \$/\mathrm{p.u}$. Specifically, 
the daily operating cost decreases from $1.105\times10^{6}$ to $9.319\times10^{5}$ USD, yielding a 15.68\% reduction, while the average effective delivered DR capacity reaches 178.3 MW. The recovery ramping rate is 333.3 MW, reflecting the grid-side response intensity. These results show that the economic benefit and DR capability mainly arise from temporal rescheduling of key process units and price-responsive grid-interactive scheduling. \par
\begin{figure}[htbp]
    \centering
    \vspace{-2mm}
    \includegraphics[width=1.0\columnwidth,trim=0 3mm 0 0,clip]{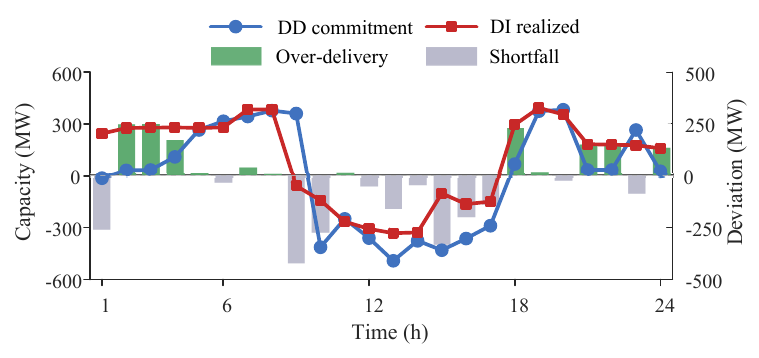}
    \vspace{-2mm}
    \caption{Day-ahead commitment, intra-day realization, and delivery deviations.}
    \label{fig:da_id_deviation}
    \vspace{-3mm}
\end{figure}
The proposed process-aware DR model effectively converts day-ahead offers into deliverable intra-day responses. As shown in Fig. 4, the intra-day realized capacity generally tracks the day-ahead commitment in both upward and downward regulation directions. Over-delivery is observed at 2:00-3:00 and 18:00, mainly because updated RES and price information lead to a more aggressive regulation response than the day-ahead commitment. In contrast, the main shortfalls at 9:00-10:00 and 15:00 reflect economic trade-offs among electricity-selling profits, shortfall penalties, and production-order constraints. Overall, the offer-realization consistency confirms that the proposed model balances real-time economic operation with physical executability.\par
The proposed process-aware DR model significantly improves the temporal coordination between RES generation and industrial demand, raising the RES-load matching degree from 0.257 to 0.587. As illustrated in Fig. 5, the AE, SF, EAF, and MSR are jointly rescheduled in response to updated RES availability. In particular, during the high RES generation period from 8:00 to 16:00, these major units increase their electricity consumption and material throughput. Consequently, their unit-level adjustment actions are made physically observable and measurable through NRIs of 2.38, 2.03, 1.74, and 0.82 for the AE, SF, EAF, and MSR, respectively. These results confirm that the overall DR performance is rooted in synergistic operational flexibility across the interconnected hydrogen-iron-steel-chemical production chain.\par
\begin{figure}[htbp]  
\includegraphics[width=1\columnwidth]{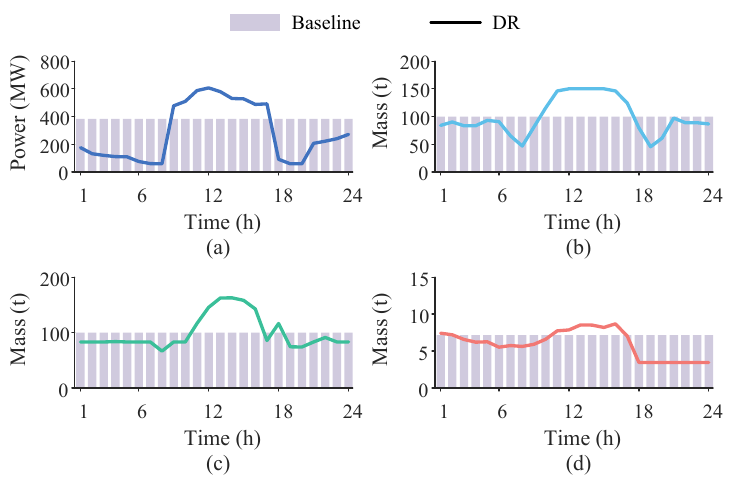}
    \caption{ Operating trajectories of key process units under baseline and DR scenarios: (a) AE, (b) SF, (c) EAF, and (d) MSR.}
    \vspace{-0.25cm}
\end{figure}
Furthermore, the proposed DR strategy enhances operating economy by reshaping the system's electricity procurement pattern. As shown in Fig. 6, the proposed DR model achieves bi-directional energy arbitrage. During price peaks (7:00–8:00 and 19:00–20:00), the proposed model replaces power purchases with active electricity exports to maximize revenues. Conversely, during low-price valleys (12:00–16:00), the proposed model replaces BI’s power exports with active purchases to exploit low-cost electricity. Additionally, the peak purchased power is reduced from 279.5 MW (BI) to 107.5 MW (DI), representing a 61.54\% reduction. Fundamentally, this confirms that the proposed DR strategy can transform conventionally rigid industrial processes into highly flexible market resources, hedging against both energy market volatility and peak capacity penalties. \par
\begin{figure}[htbp]
    \centering 
    \includegraphics[width=1\columnwidth]{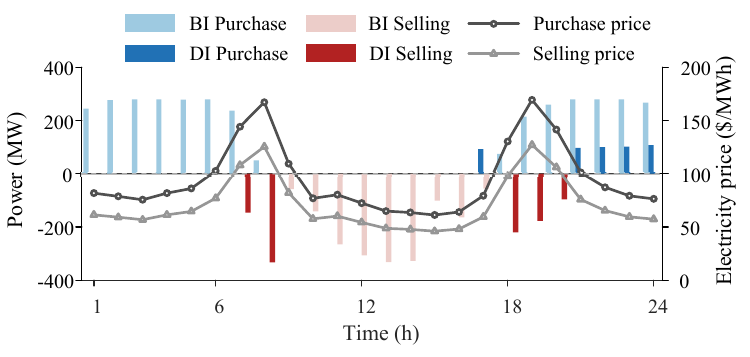}
    \caption{Price-responsive grid-interactive operation under BI and DI scheduling.}
    \label{fig:scheduling_profiles}
\end{figure}
\vspace{-0.75cm}
\subsection{Validation of the adaptive rolling mechanism}
To validate the effectiveness of the adaptive rolling mechanism (ARM), a future-capacity feasibility bound (FCFB) is used as the benchmark. The FCFB only requires the residual order after the current rolling window to remain feasible under the maximum production capacity of the remaining periods. \par
Compared with FCFB, the proposed ARM significantly improves both economic performance and dynamic source-load coordination.  Specifically, the total operating cost is reduced by 6.05\% (from $9.919\times10^{5}$ to $9.319\times10^{5}$ USD), and the RES-load matching degree increases from 0.505 to 0.587. As shown in Fig. 7(a) and (c), the ARM tracks cumulative progress to front-load production tasks, thereby preventing end-of-horizon catch-up. This physical front-loading directly translates into substantial shifts in unit-level energy flows, as illustrated in Fig. 7(b) and (d). Ultimately, the proposed ARM effectively mitigates MPC myopia, allowing the system to better coordinate order fulfillment with RES availability and electricity-price variations.\par
\begin{figure}[t!] 
\centerline{\includegraphics[width=1\columnwidth]{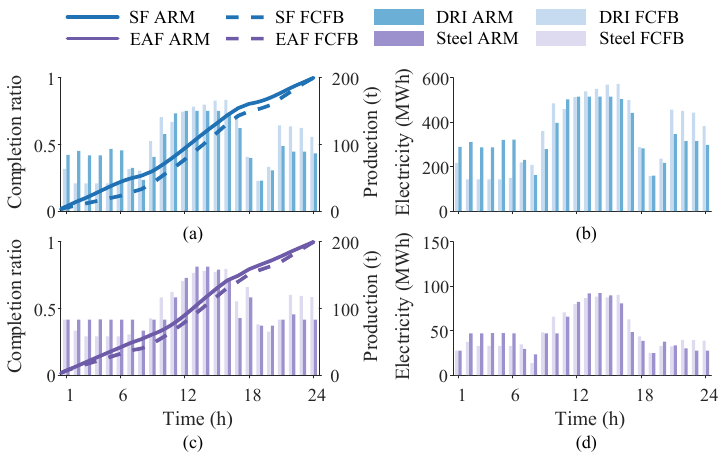}}
    \caption{Comparative analysis of order fulfillment and energy flows under ARM and FCFB: (a) SF completion ratio and production; (b) SF electricity consumption; (c) EAF completion ratio and production; (d) EAF electricity consumption.}
    \vspace{-0.5cm}
\end{figure}
Additionally, the proposed ARM effectively mitigates RES curtailment and substantially reduces grid power purchases. 
As shown in Fig. 8, the FCFB strategy curtails 530.3 MWh of RES during 2:00 to 6:00. In contrast, the ARM nearly eliminates RES curtailment and reduces grid purchases during 17:00–24:00, with cumulative purchases decreasing from 747.1 to 405.5 MWh and the peak hourly import decreasing from 192.9 to 107.5 MW, which significantly alleviates the peaking pressure on the main grid, thereby enhancing the overall grid-friendliness of the system.
\begin{figure}[t!]
\centering
\includegraphics[width=\columnwidth]{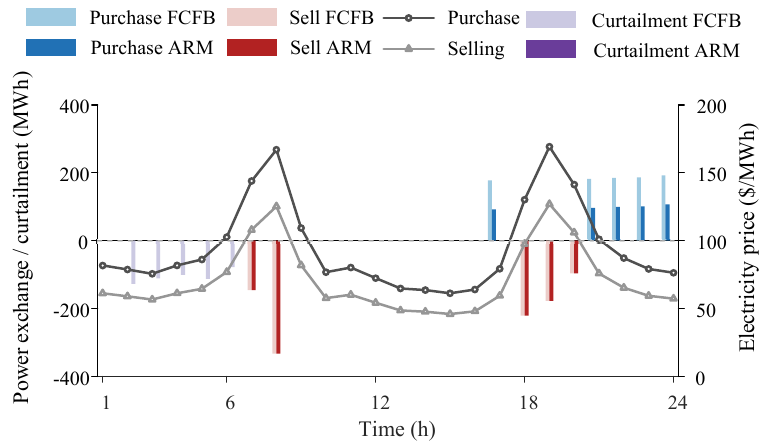}
\caption{System-level grid power trading profiles and bi-directional energy arbitrage behaviors under ARM and FCFB strategies.}
\vspace{-0.45cm}
\end{figure}
\subsection{Performance Evaluation of the Process-deviation Penalty Model: Cost Trade-off and Mechanism Comparison}
The sensitivity analysis reveals a clear nonlinear trade-off between total economic cost and the process-state deviation rate under varying penalty coefficients. As shown in Fig. 9, when the penalty coefficient increases from 0 to 100, the process-state deviation rate drops sharply from approximately 18.28 to 2.63, while the total economic cost remains within a relatively low range and reaches its minimum of $9.319\times10^{5}$ USD at a penalty coefficient of 50 \$/p.u. These results indicate that moderate penalty coefficients can correct excessive process deviations while maintaining, or even improving, overall economic performance. When the penalty coefficient increases from 200 to 10000, the process-state deviation rate only decreases marginally, whereas the total economic cost rises continuously from about $1.011\times10^{6}$ to $1.124\times10^{6}$ USD. This indicates that excessive penalties mainly restrict scheduling flexibility, yielding limited stability gains at a substantially higher economic cost. When the penalty coefficient increases from 10000 to 20000, the deviation rate approaches zero and the total economic cost declines from its peak, yielding an average delivered DR capacity of 69.0 MW. This indicates that while the system retains its DR potential, the extreme penalty completely locks down internal equipment flexibility, forcing the system to provide DR under a rigid operation mode.
\begin{figure}[t!]   
\centerline{\includegraphics[width=1\columnwidth]{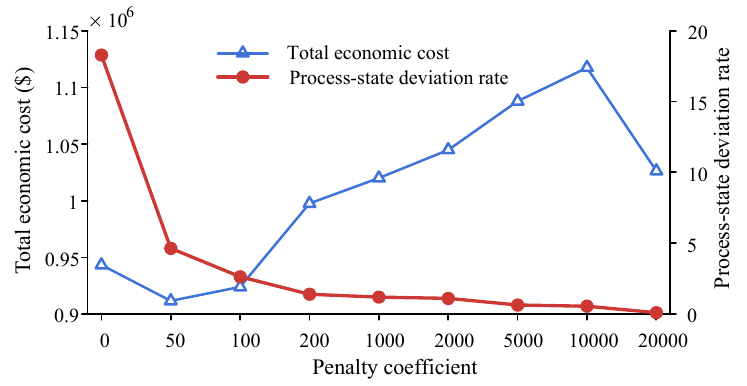}}
    \caption{Sensitivity of operating cost and process-state deviation rate to the varying penalty coefficients.}
    \vspace{-0.5cm}
\end{figure} 
\par
To demonstrate the ability of the proposed nonlinear asymmetric penalty model to suppress tail risks in operating-state deviations, three process-aware scheduling mechanisms are compared. \textbf{M1} adopts a linear symmetric penalty without a deadband, \textbf{M2} adopts a non-exponential asymmetric penalty with a deadband, and \textbf{M3} adopts the proposed exponential asymmetric penalty with a deadband. \par
Fig. 10 shows that the proposed M3 mechanism substantially reduces the tail risks of normalized process deviations compared with \textbf{M1} and \textbf{M2}. For the SF, \textbf{M1} and \textbf{M2} still produce several large positive and negative deviations, indicating that linear or non-exponential penalties are insufficient to suppress extreme operating-state deviations. In contrast, \textbf{M3} significantly concentrates the deviation distribution within the deadband and eliminates most outliers, demonstrating stronger suppression of extreme SF process-state deviations. A similar pattern is observed for the EAF: compared with \textbf{M1} and \textbf{M2}, \textbf{M3} narrows the deviation range and reduces the occurrence of large deviations, thereby improving process stability under DR scheduling. Additionally, \textbf{M1}, \textbf{M2}, and \textbf{M3} yield total operating costs of $9.285\times10^{5}$, $9.466\times10^{5}$, and $9.319\times10^{5}$ USD, respectively. Although \textbf{M3} slightly increases the cost compared with \textbf{M1}, it remains lower than \textbf{M2} while substantially reducing extreme process deviations. These results confirm that the proposed exponential asymmetric penalty formulation is effective in constraining extreme process deviations, which is consistent with operators’ aversion to severe process disturbances, without causing a substantial loss of economic performance.
\begin{figure}[t!]   
\centerline{\includegraphics[width=1\columnwidth]{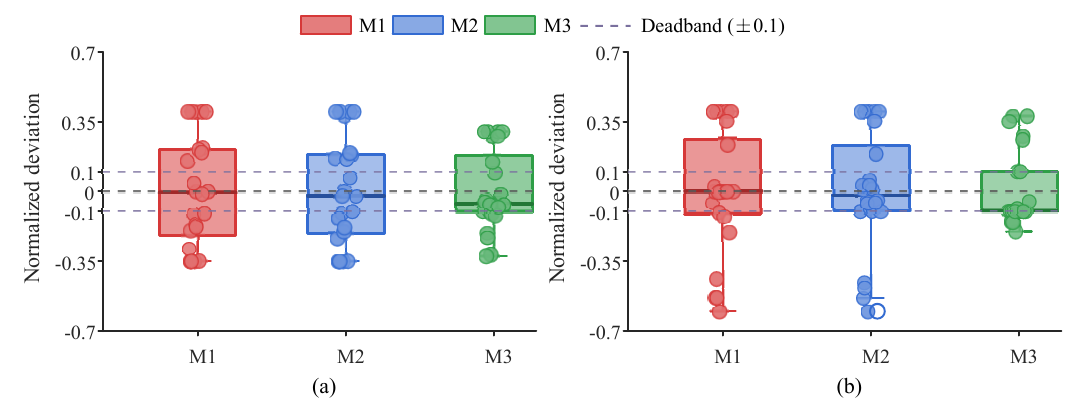}}
    \caption{Distributional comparison of the normalized deviation under different process-aware scheduling mechanisms: (a) SF; (b) EAF.}
    \vspace{-0.5cm}
\end{figure} 
\section{Conclusion}\label{Conclusion}
For the proposed \ce{H2}-DRI-EAF-MeOH system, this paper develops an EAF operating feasible region model, proposes a process-aware DR potential evaluation framework, and establishes grid-load evaluation metrics. The main conclusions are as follows: \par
1) The proposed system exhibits substantial grid-interactive flexibility under real-time electricity pricing. Compared with the BI scheme, the daily operating cost decreases from $1.105\times10^{6}$ to $9.319\times10^{5}$ USD, corresponding to a 15.68\% reduction, while the average effective delivered DR capacity reaches 178.3 MW. \par
2) The proposed ARM effectively mitigates MPC myopia by dynamically coordinating order fulfillment with RES availability and electricity-price variations. Compared with FCFB, ARM actively improves source-load coordination by reducing renewable curtailment, ultimately reducing the total operating cost by 6.05\%. \par
3) The proposed nonlinear asymmetric process-aware penalty model effectively suppresses extreme tail risks of operating-state deviation without causing a substantial loss of economic performance. \par
Future work will scale this framework from a single plant to regional industrial parks, unlocking the large-scale DR potential of industrial clusters. \par
\section{APPENDIX}\label{Proof}
This appendix provides a concise proof of the recursive feasibility and terminal reachability of the proposed ARM.
\begin{equation}
R_v^k=M_v^{\mathrm{order}}-\sum_{\tau=1}^{k-1}M_v^{\tau,\mathrm{real}},
\qquad
Q_v^k=\sum_{t\in\mathcal{T}_k^L}M_v^{t,k,\mathrm{DI}},
\end{equation}
$R_v^k$ denotes the remaining order at the beginning of step $k$, and $Q_v^k$ denotes the planned production within the current prediction horizon. The proposed pacing constraints can be rewritten as
\begin{equation}
\frac{L}{T-k+1}R_v^k \le Q_v^k \le R_v^k ,
\end{equation}
First, recursive feasibility is guaranteed if the remaining order is physically executable within the remaining scheduling period. Let $\bar{M}_v$ be the maximum production capacity per time step. Then
\begin{equation}
R_v^k \le (T-k+1)\bar{M}_v ,
\end{equation}
Therefore, the lower bound satisfies
\begin{equation}
\frac{L}{T-k+1}R_v^k \le L\bar{M}_v ,
\end{equation}
which is no larger than the maximum physical production capacity within the prediction horizon. Meanwhile, since $L\le T-k+1$, it follows that
\begin{equation}
\frac{L}{T-k+1}R_v^k \le R_v^k .
\end{equation}
Thus, the lower and upper bounds are mutually consistent, and the feasible region of $Q_v^k$ is non-empty.
Second, when the rolling horizon reaches the terminal stage, the remaining scheduling length equals the prediction horizon, i.e.,
\begin{equation}
L=T-k+1,
\end{equation}
The pacing constraints then become
\begin{equation}
R_v^k \le Q_v^k \le R_v^k ,
\end{equation}
which directly gives
\begin{equation}
Q_v^k=R_v^k .
\end{equation}
Hence, the remaining order must be exactly completed in the terminal horizon, guaranteeing terminal satisfaction.
\bibliographystyle{IEEEtran}
\bibliography{reference.bib}

\end{document}